# Structural, elastic, and electronic properties of newly discovered Li$_2$PtSi$_3$ superconductor: Effect of transition metals

M. A. Alam[1], M. A. Hadi[2], M. T. Nasir[2], M. Roknuzzaman[3], F. Parvin[2], M. A. K. Zilani[1], A. K. M. A. Islam[2,4] and S. H. Naqib[2*]

[1]Department of Physics, Rajshahi University of Engineering and Technology, Rajshahi-6204, Bangladesh
[2]Department of Physics, University of Rajshahi, Rajshahi-6205, Bangladesh
[3]Department of Physics, Jessore University of Science and Technology, Jessore-7408, Bangladesh
[4]International Islamic University Chittagong, 154/A College road, Chittagong, Bangladesh

**Abstract**

First-principles calculations within the density functional theory (DFT) with GGA-PBE exchange-correlation scheme have been employed to predict the structural, the elastic and the electronic properties of newly discovered lithium silicide superconductor, Li$_2$PtSi$_3$, for the first time. All the theoretical results are compared with those calculated recently for isostructural Li$_2$IrSi$_3$. The present study sheds light on the effect of replacement of transition metal element Ir with Pt on different mechanical, electronic, and superconducting properties. The effect of spin-orbit coupling on electronic band structure was found to be insignificant for Li$_2$PtSi$_3$. The difference in superconducting transition temperatures of Li$_2$PtSi$_3$ and Li$_2$IrSi$_3$ arises primarily due to the difference in electronic energy density of states at the Fermi level. Somewhat reduced Debye temperature in Li$_2$PtSi$_3$ plays a minor role. We have discussed the implications of the theoretical results in details in this study.

## 1. Introduction

Because of comparable electronegativity with silicon, transition metals tend to form strong covalent bond with it**.** In realizing phonon-mediated superconductivity with a high $T_c$, the covalent character of atomic bonding can play an important role. Transition metal silicides were identified as superconductors at the middle of 20$^{th}$ century. V$_3$Si has been the best example for early silicide superconductor with $T_c$ = 17 K [1]. The possibility of tuning the chemical potential (Fermi energy, in practice) due to different transition metals can, in theory, provide with a way to synthesize new silicide superconductors with significantly higher $T_c$. Due to this reason, superconductivity in metallic silicides has drawn noticeable attention. Hirai and coworkers reported about the synthesis of silicide superconductors Li$_2$IrSi$_3$, Li$_2$PtSi$_3$, and Li$_2$RhSi$_3$ in a recent publication [2]. Among these newly synthesized materials, Li$_2$IrSi$_3$ has a $T_c$ of 3.7 K, highest among the three. Li$_2$IrSi$_3$ crystallizes in the hexagonal structure with space group $P6_3/mmc$ (No. 194). The crystal structure is composed of Si triangles connected by Ir atoms, resulting in a three-dimensional network of covalent bonds. The antiprisms, IrSi$_6$ stacking along the $c$-axis, are connected by Si triangles. Li$_2$PtSi$_3$ and Li$_2$RhSi$_3$ are isostructural with Li$_2$IrSi$_3$. Theoretical studies of various physical properties of Li$_2$IrSi$_3$ have been carried out recently [3, 4] but no theoretical work has been done on Li$_2$PtSi$_3$ and Li$_2$RhSi$_3$ till date. This motivates us to study theoretically the effect of transition metal replacement in silicide superconductors. In this paper, the structural, elastic and electronic properties of Li$_2$PtSi$_3$ have been explored using the first-principles calculations and compared with Li$_2$IrSi$_3$. Replacing Ir with Pt affects all the properties under study. We have discussed these in in the subsequent sections.

The rest of the paper has been organized as follows: In Section 2, a short description of the computational methods employed here has been presented. The results are presented and discussed in Section 3. Section 4 consists of a brief summary of the findings and conclusions.

## 2. Computational Procedures

We have used hexagonal $P6_3/mmc$ (space group No. 194), to determine the structural parameters of Li$_2$PtSi$_3$ via geometry optimization. The first-principles calculations were performed by using the Cambridge Serial Total Energy Package (CASTEP) code [5], which is based on the widely employed

---

density functional theory (DFT) [6, 7] within the plane-wave pseudopotential approach. The Kohn-Sham equations were solved using the Perdew-Burke-Ernzorhof generalized gradient approximation (PBE-GGA) [8] for the exchange-correlation. Vanderbilt-type ultrasoft pseudopotentials [9] were applied to model the electron-ion interactions. Throughout the calculations, a plane-wave cutoff energy of 500 eV was chosen to determine the number of plane waves in the expansion. The crystal structures were fully relaxed by the Broyden-Fletcher-Goldfrab-Shanno (BFGS) optimization algorithm [10]. Special $k$-points sampling integration over the Brillouin zone was performed by using the Monkhorst–Pack method [11] with a 11×11×6 mesh. The tolerance of the various parameters are as follows – the energy tolerance used in the ab-initio calculations is $5×10^{-7}$ eV/atom, the magnitude of change in total energy is $5×10^{-6}$ eV/atom, the maximum force is 0.01 eV/Å, the maximum stress is 0.02 GPa, and that for the maximum atomic displacement is $5×10^{-4}$ Å. Spin-orbit coupling (SOC) was also incorporated for electronic band structure calculations and geometry optimization. For the silicide under study, SOC does not have a significant effect on various physical properties.

CASTEP calculates the elastic properties from the first-principles using the finite strain theory [12], which gives the elastic constants as the proportionality coefficients relating the theoretical strain ($\varepsilon_j$) to the computed stress, $\sigma_i = C_{ij}\,\varepsilon_j$. From the values of $C_{ij}$, the polycrystalline bulk modulus $B$ and shear modulus $G$ can be evaluated using the Voigt-Reuss-Hill approximation [13-15]. In addition, the Young's modulus $Y$, Poisson's ratio $v$, and shear anisotropy factor $A$, can be estimated using the equations $Y = (9GB)/(3B + G)$, $v = (3B – 2G)/(6B + 2G)$ and $A = 4C_{44}/(C_{11} + C_{33} – 2C_{13})$, respectively. The Debye temperature, $\theta_D$, is determined from the following equations [16]: $\theta_D = h/k_B [3n/(4\pi V)]^{1/3} v_m$, with $v_m = [1/3(1/v_l^3 + 2/v_t^3)]^{-1/3}$, $v_l = [(3B + 4G)/3\rho]^{1/2}$, and $v_t = [G/\rho]^{1/2}$. Here, $v_l$ and $v_t$ are the longitudinal and transverse sound velocities which determine the average sound velocity, $v_m$, in crystalline solids. $V$ is the unit cell volume and $n$ being the number of atoms within a unit cell.

The theoretical estimates of Vickers' hardness for crystals with metallic bonding is evaluated from the well established empirical formula [17, 18]: $H_V = [\prod^{\mu}(H_v^{\mu})^{n^{\mu}}]^{1/\Sigma n^{\mu}}$, with the individual bond hardness given by, $H_v^{\mu} = 740(P^{\mu} - P^{\mu'})(v_b^{\mu})^{-5/3}$. Here $P^{\mu}$ is the Mulliken population of the $\mu$-type bond, $P^{\mu} = n_{free}/V$ is the metallic population, and $v_b^{\mu}$ is the volume of $\mu$-type bond.

It should be noted that the elastic properties are closely related to the crystal structure and the nature of bonding among the ions within a compound. These factors, in turn, determine the phonon spectrum and the Debye temperature. Therefore, elastic constants are not only the primary parameters for the understanding of mechanical properties; they have relevance to the phenomenon of superconductivity in superconducting compounds as well.

## 3. Results and discussion

### 3.1. Structural properties

The ground state structure of $Li_2PtSi_3$ optimized with respect to lattice constants and internal atomic positions in the present investigation is shown in Figure.1. The unit cell contains two formula units consisting of 12 atoms. The optimized Li atoms are situated at the 4$f$ Wyckoff site with fractional coordinates (1/3, 2/3, 0.5663). The Si atoms occupy the 6$h$ Wyckoff position with fractional coordinates (0.3527, 0.17634, 3/4). The Pt atoms are positioned at the 2$a$ Wyckoff site with fractional coordinates (0, 0, 0). The calculated results for the structural properties of $Li_2PtSi_3$ are presented in Table 1 along with the available data for the isostructural $Li_2IrSi_3$. It can be seen from Table 1 that the replacement of Ir with Pt affects the lattice constants, unit cell volume, and bulk modulus. The lattice constants $a$ and $c$ increase by 1.30% and 0.25%, respectively, whereas, the unit cell volume $V$, increases by 2.87% compared to the experimental value of that of $Li_2IrSi_3$ [2]. The bulk modulus $B$ decreases significantly, by 8.40%, with respect to the theoretical result of $Li_2IrSi_3$ [4].

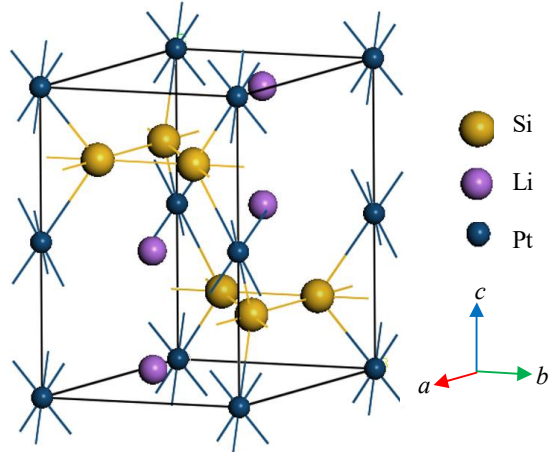

**Fig. 1.** The unit cell of hexagonal silicide superconductor $Li_2PtSi_3$.

**Table 1**. Lattice constants $a$ and $c$ (in Å), internal parameter $z_{Li}$, hexagonal ratio $c/a$, unit cell volume $V$ (in Å$^3$), and bulk modulus $B$ (GPa) with its pressure derivative $B'$ for hexagonal $Li_2PtSi_3$.

| Compounds | $a$ | $c$ | $c/a$ | $z_{Li}$ | $V$ | $B$ | $B'$ | Remarks |
|---|---|---|---|---|---|---|---|---|
| $Li_2PtSi_3$ | 5.0826 | 7.8600 | 1.5465 | 0.5663 | 175.84 | 106.16 | 4.50 | This study |
| $Li_2IrSi_3$ | 5.0176 | 7.8402 | 1.5625 | 0.5604 | 170.94 | | | Expt. [2] |
| | 5.0069 | 7.8655 | 1.5709 | 0.5589 | 170.76 | 115.90 | 4.24 | Calc. [4] |
| | 5.0183 | 7.8414 | 1.5626 | | 170.42 | | | Calc. [3] |

### 3.2. Elastic properties

Calculated elastic coefficients of $Li_2PtSi_3$ are listed in Table 2 along with the previously calculated values for isostructural $Li_2IrSi_3$ [4]. The positive values of $C_{11}$, $C_{44}$, $(C_{11} - C_{12})$, and $\{(C_{11} + C_{12})C_{33} - 2C_{12}^2\}$ conform with the Born mechanical stability criteria [19] for $Li_2PtSi_3$. The small values of $C_{12}$ and $C_{13}$ suggest that $Li_2PtSi_3$ should be classified as a brittle material [20]. The widely used Pugh's criterion [21] and the Frantsevich's rule [22] also favor this classification. Pugh's criterion suggests that a material will be brittle if its $G/B > 0.5$, otherwise it is ductile. The Frantsevich's rule states that the mechanical property of a material will be dominated with brittleness when Poisson's ratio $v < 0.33$; if $v > 0.33$, the mechanical property of the material mainly exhibits ductile nature. As can be seen from Table 2, the newly synthesized $Li_2PtSi_3$ should be brittle in nature like $Li_2IrSi_3$. Qualitatively, substitution of Ir with Pt reduces the brittleness of these compounds. In fact, $Li_2PtSi_3$ lies at the borderline between ductility and brittleness. The comparatively high value of Poisson's ratio indicates that the directional covalent bonding in $Li_2PtSi_3$ is not as strong as in $Li_2IrSi_3$.

The difference between $C_{11}$ and $C_{33}$ indicates about the elastic anisotropy possessed by the crystal. Elastic anisotropy results from different types of bonding in different directions. Inherently, majority of the known crystals are elastically anisotropic, and an accurate description of such anisotropic behavior has a significant implication in engineering, applied science, and crystal physics, since it correlates with the possibility of appearance of micro-cracks inside the crystals under stress. It is seen that the difference between $C_{11}$ and $C_{33}$ reduces when Ir is substituted by Pt. Therefore, the elastic anisotropy in $Li_2PtSi_3$ should be low in comparison with $Li_2IrSi_3$. We have also estimated the shear anisotropy factor, given by $A = 4C_{44}/(C_{11} + C_{33} - 2C_{13})$. For an isotropic crystal, $A = 1$, and the deviation from unity measures the degree of elastic anisotropy. The calculated results show that the phase $Li_2PtSi_3$ is almost isotropic, whereas $Li_2IrSi_3$ possesses small anisotropy for the shear plane $\{100\}$ between the directions $\langle 011 \rangle$ and $\langle 010 \rangle$. Yet, another anisotropy parameter can be obtained from the ratio between the uniaxial compressions along the $c$- and $a$-axis for a hexagonal crystal: $k_c/k_a$

$= (C_{11} + C_{12} - 2C_{13})/(C_{33} - C_{13})$. The deviation from unity for $Li_2PtSi_3$ is small in comparison with $Li_2IrSi_3$ and it lacks elastic anisotropy [23].

The ratio between the bulk modulus $B$ and elastic constant $C_{44}$ may be interpreted as a gauge of plasticity [24]. The plasticity can also be estimated from the value of $(C_{11} - C_{12})$ and the Young's modulus $Y$ [25]. The smaller values of $(C_{11} - C_{12})$ and $Y$ indicate better plasticity of $Li_2PtSi_3$ in comparison with $Li_2IrSi_3$. Furthermore, $B/C_{44}$ ratio measures the lubricity which gives the index of machinability of bulk materials. The values of this ratio signify that the degree of lubricity and machinability of $Li_2PtSi_3$ is almost similar to $Li_2IrSi_3$. Bulk and shear moduli are considered as an important parameter for estimating the hardness of bulk materials. The small values of these two parameters for $Li_2IrSi_3$ and $Li_2PtSi_3$ imply that they may be grouped into high compressible materials. Again, in comparison to $Li_2IrSi_3$, $Li_2PtSi_3$ is more compressible. Young's modulus gives a measure of stiffness and it is seen that substitution of Pt in place of Ir makes the compound less stiff (Table 2). Therefore, Pt causes a significant reduction of the independent elastic constants ($C_{11}$, $C_{33}$, and $C_{44}$) and all the elastic moduli ($B$, $G$, and $Y$). At the same time, this substitution results in a lowering of the brittleness, elastic anisotropy, and strength of the directional covalent bonding in $Li_2PtSi_3$.

**Table 2**. Single crystal elastic constants $C_{ij}$ (GPa), polycrystalline bulk modulus $B$ (GPa), shear modulus $G$ (GPa), Young modulus $Y$ (GPa), Pugh's ratio $G/B$, Poisson's ratio $v$, and anisotropy factor $A$ and $k_c/k_a$ of $Li_2PtSi_3$ and $Li_2IrSi_3$.

| Compounds | Single crystal elastic constants | | | | | Polycrystalline elastic properties | | | | | | | Remarks |
|---|---|---|---|---|---|---|---|---|---|---|---|---|---|
| | $C_{11}$ | $C_{12}$ | $C_{13}$ | $C_{33}$ | $C_{44}$ | $B$ | $G$ | $Y$ | $G/B$ | $v$ | $A$ | $k_c/k_a$ | |
| $Li_2PtSi_3$ | 160 | 82 | 63 | 224 | 67 | 106 | 55 | 141 | 0.52 | 0.28 | 1.04 | 0.72 | This study |
| $Li_2IrSi_3$ | 198 | 71 | 55 | 295 | 68 | 116 | 72 | 179 | 0.62 | 0.24 | 0.71 | 0.66 | Calc. [4] |

The Debye temperature, calculated from longitudinal, transverse, and average sound velocities within the present formalism is listed in Table 3. It is observed that the replacement of Ir with Pt results in an appreciable reduction in the Debye temperature. It is known that the Debye temperature is indicative of the strength of covalent bonding in a crystal. Therefore, we may conclude again that the covalent bonding in $Li_2PtSi_3$ is weaker than that of $Li_2IrSi_3$.

**Table 3**. Calculated density ($\rho$ in gm/cm$^3$), longitudinal, transverse, and average sound velocities ($v_l$, $v_t$, and $v_m$ in km/s) and Debye temperature ($\theta_D$ in K).

| Compounds | $\rho$ | $v_l$ | $v_t$ | $v_m$ | $\theta_D$ |
|---|---|---|---|---|---|
| $Li_2PtSi_3$ | 5.538 | 5.691 | 3.151 | 3.510 | 427 |
| $Li_2IrSi_3$ | 5.641 | 6.130 | 3.573 | 3.963 | 488 [4], 486[a], 488[b] |

[a]Ref. [2][Expt.]
[b]Ref. [3][Calc.]

### 3.3. Electronic and bonding properties

In Fig. 2a, the energy band, without SOC, of $Li_2PtSi_3$ with optimized lattice parameters along the high symmetry directions within the first Brillouin zone is presented. The Fermi level lies just below the lowest conduction band near the Γ point. The occupied valence bands spread from –12.8 eV to the Fermi level $E_F$, and few of them just crosses the Fermi level along the M-L direction. Further, two of valence bands overlap with the conduction band near the M point. Consequently, no band gap is seen and the compound behaves as a metallic system like the isostructural $Li_2IrSi_3$. The gross features of the band structures of $Li_2PtSi_3$ and $Li_2IrSi_3$ are almost similar.

The total and partial energy density of states (DOS) of $Li_2PtSi_3$ is shown in Fig. 2b. The total DOS has a finite value at the Fermi level $N(E_F)$, around 1.48 states per eV per unit cell, which is small in comparison with the values found in literatures for $Li_2IrSi_3$ [2-4] (~ 3.5 states per eV per unit cell). The DOS at the Fermi level reduces significantly due to the replacement of Ir by Pt. Since the electron phonon coupling constant, $\lambda$, varies linearly with $N(E_F)$, a lower value of electronic density of states at the Fermi level implies a lower value of $T_c$, given that the matrix element of electron-phonon interaction is fixed. This provides us with an explanation for the reduced value of $T_c = 1.8$ K for $Li_2PtSi_3$ in comparison to that for $Li_2IrSi_3$, where $T_c = 3.7$ K.

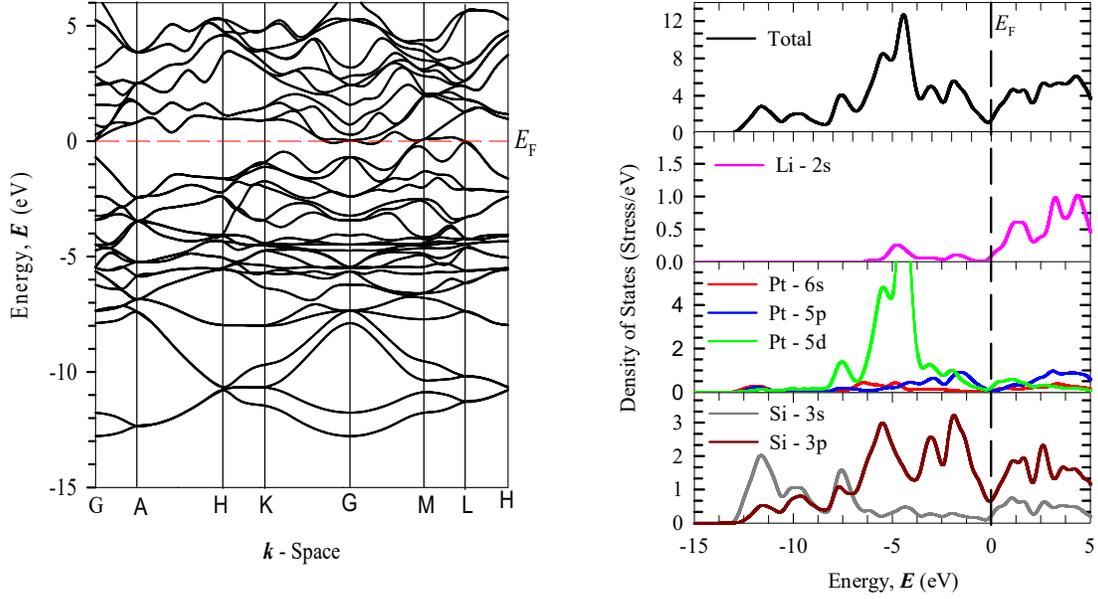

**Fig. 2.** Electronic structure of $Li_2PtSi_3$: (a) Band structure at equilibrium lattice parameters along the high symmetry directions within the Brillouin zone and (b) Total and partial energy density of states. The Fermi level is set to 0 eV.

The details of the peak structures and the relative heights in the total and partial DOSs for $Li_2PtSi_3$ are roughly similar to those found for $Li_2IrSi_3$ [3, 4]. One notable difference is that the DOS at the Fermi level for $Li_2PtSi_3$ is situated at a dip and a pseudogap appeared at the left of the Fermi level, whereas the DOS at the Fermi level for $Li_2IrSi_3$ lies just below a peak structure.

Inspection of the partial DOSs reveals that the DOS at the Fermi level originates mainly from Pt $5d$ and Si $3p$ states. This is due to the $d$-resonance in the vicinity at the Fermi level, similar to the case for $Li_2IrSi_3$ where the $d$ electrons of the Ir atoms play a similar role. The lowest-lying valence bands of both these silicide superconductors with two peak structures consisting of $3s$ and $3p$ states of Si occupy almost same energy range. But this band for $Li_2PtSi_3$ is not as wide as $Li_2IrSi_3$ [4]. The covalent Si–Si bond that forms a three dimensional rigid network as a whole in case of $Li_2IrSi_3$ [2] is stronger than that of $Li_2PtSi_3$. The rest of the valence band arises mainly from the strong hybridization of $3s$ and $3p$ states of silicon with $5d$ states of transition metal Pt in $Li_2PtSi_3$ or Ir in $Li_2IrSi_3$.

**Table 4**. Atomic population analysis of $Li_2PtSi_3$ and $Li_2IrSi_3$.

| Compounds | Species | Mulliken Atomic populations | | | | | Effective valence Charge (e) |
|---|---|---|---|---|---|---|---|
| | | s | p | D | Total | Charge (e) | |
| $Li_2PtSi_3$ | Li | 1.90 | 0.00 | 0.00 | 1.90 | 1.10 | 1.90 |
| | Pt | 0.91 | 1.62 | 9.02 | 11.54 | − 1.54 | 3.54 |
| | Si | 1.36 | 2.86 | 0.00 | 4.22 | − 0.22 | |
| $Li_2IrSi_3$ | Li | 1.99 | 0.00 | 0.00 | 1.99 | 1.01 | 1.99 |
| | Ir | 0.77 | 1.31 | 8.26 | 10.33 | − 1.33 | 5.33 |
| | Si | 1.37 | 2.86 | 0.00 | 4.23 | − 0.23 | |

**Table 5.** Calculated Mulliken bond number $n^\mu$, bond length $d^\mu$, bond overlap population $P^\mu$, bond volume $v_b^\mu$ and bond hardness $H_v^\mu$ of $\mu$-type bond and metallic population $P^{\mu'}$ and Vickers hardness $H_v$ of Li$_2$PtSi$_3$ and Li$_2$IrSi$_3$

| Compounds | Bond | $n^\mu$ | $d^\mu$ (Å) | $P^\mu$ | $P^{\mu'}$ | $v_b^\mu$ (Å$^3$) | $H_v^\mu$ (GPa) | $H_v$ (GPa) |
|---|---|---|---|---|---|---|---|---|
| Li$_2$PtSi$_3$ | Si–Si | 6 | 2.39337 | 0.51 | 0.00125 | 7.55875 | 12.93162 | 5.87 |
|  | Si–Pt | 12 | 2.50435 | 0.26 | 0.00125 | 8.65975 | 5.24429 |  |
| Li$_2$IrSi$_3$ | Si–Si | 6 | 2.43154 | 0.56 | 0.03573 | 9.22726 | 9.55723 | 6.21 |
|  | Si–Ir | 12 | 2.46525 | 0.33 | 0.03573 | 9.61637 | 5.00756 |  |

To explore the bonding nature of newly synthesized silicide superconductors Li$_2$PtSi$_3$ and Li$_2$IrSi$_3$, the Mulliken atomic populations are calculated for the first time. The atomic populations are listed in Table 4 along with the effective valence. The difference between the formal ionic charge and the Mulliken charge on the anion species in the crystal is defined as the effective valence that identifies a bond either as covalent or ionic. An ideal ionic bond occurs when the effective valence is exactly zero. The strength of the covalent bond increases if the effective valence shows an increasing tendency towards positive values. The last column of Table 4 represents the calculated effective valence that indicates significant covalent bonding in both the silicides. Table 5 lists the bond overlap populations for the Si-Si and Si-Pt bonds of Li$_2$PtSi$_3$ and Si-Si and Si-Ir bonds of Li$_2$IrSi$_3$. A value of the overlap population close to zero implies that the interaction between the electronic populations of the two atoms is negligible. A low overlap population also indicates a high degree of ionicity, whereas a high value implies a high degree of covalency in the chemical bond. Therefore, it follows from Table 5 that Si-Si bond is more covalent in Li$_2$IrSi$_3$ than that in Li$_2$PtSi$_3$.

The degree of metallicity may be calculated from the parameter $f_m = P^{\mu'}/P^\mu$ [16, 26]. The calculated metallicity of the Si–Si and Si-Pt bonds in Li$_2$PtSi$_3$ are 0.00245 and 0.00481, respectively. For Li$_2$IrSi$_3$, the calculated metallicity for Si–Si and Si–Ir bonds are 0.06380 and 0.10827, respectively, suggesting that the Si-Ir bond has strong metallic characteristics. Table 5 shows that the bonding nature in the two new silicide superconductors may be described as a combination of covalent and metallic. Table 5 also lists the calculated Vickers hardness that reflects the crystal structure and chemical bonding of the compounds. The total Vickers hardness is obtained by calculating the geometrical average of the individual bond hardness of the compound. The general relationship between the bond-length and hardness is such that the hardness increases with decreasing the bond-length. Considering this relationship, the Si–Si bonds are expected to be harder than other bonds. The calculated values of the Vickers hardness for Li$_2$PtSi$_3$ and Li$_2$IrSi$_3$ are found to be 5.87 and 6.21 GPa, respectively. It is evident that the Vickers hardness reduces when Ir is replaced with Pt and therefore, Li$_2$PtSi$_3$ is relatively soft and easily mechinable compared to Li$_2$IrSi$_3$.

## 3. Conclusions

A first-principles study on the structural, elastic, bonding, and the electronic properties of Li$_2$PtSi$_3$ is done for the first time. The theoretical results have been compared with those found for isostructural Li$_2$IrSi$_3$ to explore the role of Pt in place of Ir. Replacing Ir with Pt affects the lattice parameters, elastic constants, bulk modulus, shear modulus, Young's modulus, Poisson's ratio, Pugh's ratio, elastic anisotropy, Debye temperature, and Vickers hardness to a varying degree. The electronic properties such as band structure, total and partial DOS and Mulliken bond populations are also affected. The most significant change is seen in the drastically reduced electronic density of states at the Fermi level of Li$_2$PtSi$_3$. This should be the primary reason for significantly reduced superconducting $T_c$ in this compound. Using the theoretically estimated Debye temperature and the $T_c$ equation due to Allen-Dynes [27], we have calculated the electron-phonon coupling constant, $\lambda \sim 0.4$, to get the experimental $T_c = 1.8$ K. Therefore, like Li$_2$IrSi$_3$, Li$_2$PtSi$_3$ is also a weakly coupled BCS superconductor. It would be interesting to synthesize Li$_2$ReSi$_3$ and Li$_2$OsSi$_3$ in future. These silicides may exhibit superconductivity at higher temperatures because of $5d^56s^2$ and $5d^66s^2$ electronic configurations, respectively, which may enhance $N(E_F)$. As mentioned earlier, inclusion of SOC does not change most of the properties in any significant way. The only noticeable effect of SOC is the lifting of degeneracy of the high-energy electronic bands ( ~ 4 – 6 eV).